\documentstyle[psfig]{caosp}
\begin{document}
\def\Hm{\langle H\rangle}
\def\Hs{(\Hsq+\Hzsq)}
\def\Ht{\Hs^{1/2}}
\def\Hsq{\langle H^2\rangle}
\def\Hzsq{\langle H_z^2\rangle}
\hauthor{S. Hubrig {\it et al.}}
\title{HgMn Stars as apparent X-ray emitters}
\author{S. Hubrig \inst{1} \and T. W. Bergh\"ofer
\inst{2}   \and G. Mathys \inst{3}}
\institute{University of Potsdam, Am Neuen Palais 10, D-14469 Potsdam, Germany
\and Space Sciences Laboratory, University of California,
    Berkeley, CA 94720-7450, USA \and 
European Southern Observatory, Casilla 19001, Santiago 19, Chile}

\date{December 16, 1997}
\maketitle
\begin{abstract}
In the ROSAT all-sky survey 11 HgMn stars were detected as soft X-ray emitters
(Bergh\"ofer, Schmitt \&\ Cassinelli 1996). Prior to ROSAT, X-ray observations
with the {\em Einstein Observatory} had suggested that stars in the spectral 
range B5--A7 are devoid of X-ray emission. Since there is no X-ray emitting 
mechanism available for these stars (also not for HgMn stars), the usual 
argument in the case of an X-ray detected star of this spectral type is the 
existence of an unseen low-mass companion which is responsible for the X-ray 
emission. The purpose of the present work is to use all available data for 
our sample of X-ray detected HgMn stars and conclude on the nature of possible 
companions. 
\end{abstract}
\section{Introduction}
\label{intr}
In the ROSAT all-sky survey X-ray emission was detected in 11 HgMn stars 
(5 spectroscopic binaries, 4 double-lined spectroscopic binaries, and 2 HgMn 
stars without available radial velocity data). For 2 of 3 spectroscopic 
binaries (SB) additional observations obtained with the ROSAT 
High-Resolution Imager (spatial resolution of $\approx$ 5 arcsec) confirmed 
X-ray sources at the position of two systems (Bergh\"ofer \&\ Schmitt 1994). 
Known visual companions could be discarded as most likely X-ray emitters.
Previous X-ray observations (by the {\em Einstein Observatory}) had suggested
that stars in the spectral range B5--A7 are devoid of X-ray emission.
Since there is no X-ray emitting mechanism available for these stars,
the usual argument in the case of an X-ray 
detection is the existence of an unseen low-mass companion which is 
responsible 
for the X-ray emission. However, this hypothesis is not easily testable.

For our sample of X-ray detected HgMn stars we used all available data to
conclude on the nature of the companion.
We emphasize that some of our sample stars consist 
of two nearly equal B stars. The observed X-ray emission in these systems is 
also inconsistent with the secondary star and, thus, a third component 
must exist to explain the X-ray emission by a low-mass companion. 
Some of our sample stars show X-ray luminosities that exceed the X-ray output 
of normal late-type stars and, therefore, an active pre-main sequence 
companion (PMS) is required. This hypothesis is supported by the fact that a 
significant fraction of the HgMn stars found in the ROSAT survey belong to 
rather young stellar groups like the Pleiades supercluster or the Sco-Cen 
association. Recently, it
has been shown that there is a population of pre-main sequence stars in the 
Pleiades supercluster, and that both cluster and non-cluster members range in 
age from about $2.6 \cdot 10^6$ to $10^8$ yrs (Eggen 1995; Oppenheimer et 
al. 1997). Many of these stars exhibit high levels of stellar activity and 
strong lithium lines.  If there is ongoing star formation in these regions, the
phenomenon demands further study and the possibility of protostars in multiple
star systems has wide-ranging implications.

\section{Results and conclusions}
Here we describe our method to conclude on the nature of possible low-mass
companions. In a first step stellar masses and ages were derived for our 
sample of HgMn stars. For this we used the stellar model grids provided by 
Schaller et al. (1992); the stellar distances were taken from the recently 
released Hipparcos catalog and the effective temperatures were compiled from 
the literature. We then assumed that the absence of a secondary in the
optical spectrum implies a mass ration of M$_1$/M$_2 \geq 1.5$ for the two 
binary components and all systems are formed coeval. A further criterion was
the saturation limit of $\log ({\rm L}_x/{\rm L}_{Bol}) \approx -3$ known for 
late-type star X-ray emission (cf. Schmitt 1997); for the observed X-ray 
luminosities this relation provides upper limits for the bolometric 
luminosities of the possible secondaries. Together with these limits for the 
companions masses, luminosities, and ages, we used the pre-main-sequence 
evolutionary tracks provided by D'Antona and Mazzitelli (1994) to limit
the range of possible companions of the 11 HgMn stars.

For all of our sample HgMn stars detected in the ROSAT all-sky survey we find
that a companion of lower mass can provide a natural explanation for the 
observed X-ray emission. In 7 cases (HD 32964, HD 33904, HD 35497, HD 75333, 
HD 110073, HD 141556, and HD 173524) the 
detected X-ray emission can be explained by a main-sequence late-type star, 
whereas for the stars HD 27295, HD 27376, HD 29589, 
and HD 221507 a PMS star is required.
Further investigations by means of radial velocity studies and high-resolution
imaging (e.g., in the near IR) are needed to detect the predicted companions.
According to the lower mass limits derived for possible 
companions in our sample of HgMn stars, spectral types are in the range
late K-M4. 

It is remarkable that in many cases 
when a spectroscopic binary with a late-B primary has a third, distant 
companion, the SB primary is a HgMn star (e.g., Hubrig \&\ Mathys 1995).

\end{document}